\documentclass[11pt]{article}

\usepackage[utf8]{inputenc}
\usepackage[T1]{fontenc}
\usepackage[english]{babel}

\usepackage[colorlinks=true, linkcolor=blue, citecolor=blue, urlcolor=blue]{hyperref}
\usepackage{url}

\usepackage{graphicx}
\usepackage{xcolor}

\usepackage{tcolorbox}
\tcbuselibrary{listings,breakable}
\usepackage{listings}

\usepackage{amsmath,amssymb,amsfonts}

\usepackage{booktabs}
\usepackage{tabularx}
\usepackage{makecell}
\usepackage{longtable}
\newcolumntype{Y}{>{\centering\arraybackslash}X}

\usepackage{floatrow}

\usepackage{comment}
\usepackage{microtype}
\usepackage{nicefrac}

\usepackage[acronym]{glossaries}
\glsdisablehyper
\newacronym{LLM}{LLM}{Large Language Model}
\newacronym{MCQA}{MCQA}{Multiple Choice Question Answering}
\newacronym{OT}{OT}{Operational Technology}
\newacronym{IT}{IT}{Information Technology}

\usepackage[numbers,sort&compress]{natbib}

\usepackage[margin=1in]{geometry}

\usepackage{authblk}

\usepackage{orcidlink}

\usepackage{caption}

\newtcolorbox{promptbox}[1]{
  breakable,
  colback=gray!10,
  colframe=black,
  boxrule=0.5pt,
  sharp corners,
  title=#1,
  fonttitle=\bfseries,
  listing only,
  listing options={
    basicstyle=\small\ttfamily,
    breaklines=true,
    keywordstyle=\color{blue}
  }
}

\providecommand{\keywords}[1]{\par\vspace{1em}\noindent\textbf{Keywords:} #1}

\title{CyberCertBench: Evaluating LLMs in Cybersecurity Certification Knowledge}

\author{Gustav Keppler\,\orcidlink{0000-0002-2323-0533}}
\author{Ghada Elbez\,\orcidlink{0000-0003-1137-1782}}
\author{Veit Hagenmeyer\,\orcidlink{0000-0002-3572-9083}}

\affil{%
  Institute for Automation and Applied Informatics (IAI),\\
  Karlsruhe Institute of Technology (KIT), Karlsruhe, Germany\\
  \texttt{\{gustav.keppler,ghada.elbez,veit.hagenmeyer\}@kit.edu}
}

\date{}

\begin{document}

\maketitle

\begin{abstract}
The rapid evolution and use of Large Language Models (LLMs) in professional workflows require an evaluation of their domain-specific knowledge against industry standards. We introduce CyberCertBench, a new suite of Multiple Choice Question Answering (MCQA) benchmarks derived from industry-recognized certifications. CyberCertBench evaluates LLM domain knowledge against the professional standards of Information Technology cybersecurity and more specialized areas such as Operational Technology and related cybersecurity standards. Concurrently, we propose and validate a novel Proposer-Verifier framework, a methodology to generate interpretable, natural-language explanations for model performance. Our evaluation shows that frontier models achieve human expert level in general networking and IT security knowledge. However, their accuracy declines in questions that require vendor-specific nuances or knowledge in formal standards, like, e.g., IEC 62443. Analysis of model scaling trend and release date demonstrates remarkable gains in parameter efficiency, while recent larger models show diminishing returns. Code and evaluation scripts are available at: \url{https://github.com/GKeppler/CyberCertBench}.
\end{abstract}


\section{Introduction}
Large language models (LLMs) have shown potential for generative and knowledge-intensive tasks that require domain knowledge, leading to rapid integration into professional workflows. Recent empirical studies demonstrate that software engineers and IT professionals increasingly rely on conversational AI tools for fact checking, code generation, debugging, and documentation tasks~\cite{Vaithilingam.2022, Barke.2023, Khojah.2024}. Similarly, researchers have described the growing adoption of LLM-based assistants in cybersecurity workflows, where professionals use them for vulnerability assessment, incident reporting, and threat-intelligence summarization~\cite{Alam.2024, Rodriguez.2025, Zhang.2025}. Comparable trends are evident across more general high-stakes technical domains such as healthcare, engineering, and infrastructure operations~\cite{Sallam.2023,  Choi.2024a}. 
This increasing dependence underscores a broader shift toward AI-mediated decision-making in professional contexts. However, it also introduces risks of LLMs "hallucinating"~\cite{Ji.2023} plausible-sounding but unfaithful or nonsensical information, which can be particularly hazardous in domains where reliability and accuracy are critical.
A growing body of research has begun to evaluate the Information Technology (IT) security capabilities of LLMs using question-answering benchmarks~\cite{Hendrycks.2021, Tihanyi.2024} and assessing task-specific skills like cyber threat intelligence analysis and vulnerability detection~\cite{Alam.2024,Rodriguez.2025, Zhang.2025}.

However, specialized knowledge—such as the procedural, vendor-specific details required to manage enterprise security hardware or the high-stakes domain of \Gls{OT} cybersecurity is unexplored. These systems form the backbone of enterprise environments and critical infrastructure, including energy grids, water treatment facilities, where security failures can have catastrophic physical consequences~\cite{Zografopoulos.2021}. The knowledge required to secure these environments includes unique protocols, real-world physical process implications, and security frameworks like ISA/IEC 62443. These are more sparsely represented in the web-scale corpora used to train foundation models, compared to general IT knowledge. Therefore, LLMs that show high capability in IT security-related benchmarks may strongly ‘hallucinate’ plausible but incorrect guidance for OT systems, creating a false sense of security for operators and engineers~\cite{xiao-wang-2021-hallucination}. The extent to which LLMs can reliably provide guidance in these specialized areas is largely unquantified.    

The present paper addresses this challenge by introducing \textbf{CyberCertBench}, a suite of multiple-choice question answering benchmarks derived from  professional certification exams that serve as a proxy for the practical and real-world knowledge. By spanning the gradient from IT to vendor-specific IT and specialized OT, our evaluation provides the first analysis of this performance hierarchy. We quantify the knowledge gaps that emerge as the required expertise becomes more specialized, posing a direct risk to the safe adoption of LLMs. Our key contributions are:

\begin{itemize}
    \item A suite of novel benchmarks for IT and OT cybersecurity knowledge of LLMs, derived from industry-recognized certification standards.
    \item A systematic analysis of a wide range of LLMs, revealing a performance gradient where accuracy is high on benchmarks testing conceptual knowledge, both in IT and OT, but degrades significantly on those requiring specialized knowledge of vendor-specific procedures and formal standards.
    \item A novel Proposer-Verifier framework to generate interpretable descriptions of question difficulties, offering a method to understand the causes of these knowledge gaps.
\end{itemize}

\section{Related Work}
A variety of specialized benchmarks are used for evaluating the cybersecurity capabilities of \glspl{LLM}. These benchmarks have evolved from broad knowledge assessments to more practical and domain-specific evaluations.

\paragraph{\textbf{Security Knowledge Benchmarks}}
Multiple-choice question benchmarks offer measurability and scalability. Besides the MMLU Computer Security subtask~\cite{Hendrycks.2021}, WMDP~\cite{Li.2024b} offers questions on biosecurity, cybersecurity, and chemical security. CyberSecEval  uses free-response questions evaluated by another LLM. Other benchmarks like SecEval~\cite{li2023seceval}, SecQA~\cite{Liu.2023}, and CyberMetric~\cite{Tihanyi.2024} use LLMs to generate questions from textbooks and other technical documents. While comprehensive, these benchmarks often suffer from saturation, where top models achieve near-perfect scores, limiting their ability to differentiate between frontier models.  Specialized benchmarks such as CTIBench\cite{Alam.2024}, which is designed specifically for Cyber Threat Intelligence tasks, and OCCULT\cite{Kouremetis.2025}, a multiple-choice benchmark focused on offensive cybersecurity tactics, provide targeted evaluations for distinct areas within the cybersecurity domain.

\paragraph{\textbf{Agent-based and Offensive Capability Evaluation}}
Recent literature is evaluating the practical, and often offensive, capabilities of LLM-based agents. This is motivated by the need to understand potential misuse and to develop more robust defenses~\cite{Phuong.2024, Rodriguez.2025}. Early work demonstrated that LLMs could assist Pentesting PentestGPT~\cite{Deng.2023} and solving Capture the Flag Challenges~\cite{Shao.2024}, leading to the creation of dedicated benchmark datasets like the NYU CTF Dataset~\cite{Shao.2024a}. This also includes the development of sophisticated LLM agents such as AutoAttacker~\cite{Xu.2024a}, and EnIGMA~\cite{Abramovich.2024a}, and multi-agent frameworks like D-CIPHER~\cite{Udeshi.2025} designed to autonomously solve these challenges.

\paragraph{\textbf{Multitask Frameworks}}
To standardize these practical evaluations, several comprehensive frameworks have been proposed. The Catastrophic Cyber Capabilities Benchmark (3CB)~\cite{Anurin.2024} specifically focuses on assessing offensive capabilities like vulnerability exploitation and privilege escalation. CyberBench~\cite{Liu.} evaluates multiple tasks, including text classification and summarization. Concurrently, major AI labs have released their own evaluation suites, such as CYBERSECEVAL~\cite{Wan.2024,Bhatt.} by Meta or Google Deepmind~\cite{Rodriguez.2025}, which measure both the beneficial capabilities and potential risks of frontier models. These agent-based evaluations test knowledge, tool use, and planning in dynamic environments.

\paragraph{\textbf{Evaluation Using Professional Certifications}}
To ensure authenticity and quality while avoiding the synthetic nature of benchmarks, some studies have turned to professional certification exams. Tann et al.~\cite{Tann.2023} first demonstrated that \glspl{LLM} could pass Cisco certification exams. Earlier work~\cite{Keppler.2025}
also evaluated models on publicly available Cisco CCNA and CCNP materials, establishing a reproducible baseline for \gls{IT} security knowledge. 
\section{Methodology}
The methodology is centered around a novel collection of benchmarks derived from professional certifications, and compared to established computer security benchmarks. This section details the creation and composition of these benchmarks, the selection of models for evaluation, and the experimental setup.

\subsection{Benchmark Collection and Curation}
To comprehensively evaluate the cybersecurity knowledge of \glspl{LLM}, we curated a suite of \gls{MCQA} benchmarks. This collection includes established academic benchmarks~\cite{Hendrycks.2021,Tihanyi.2024} to establish a baseline, alongside several novel datasets derived from industry career certifications. This approach allows for an analysis of both general and specialized knowledge, with a particular focus on the under-evaluated domain of \gls{OT}.

\begin{table}[htbp]
  \caption{Analysis of the CyberCertBench Suite across a gradient of knowledge specificity.}
  \label{tab:cert_benchmark_analysis}
  \centering
  \begin{tabularx}{\textwidth}{p{2cm} Y Y Y Y}
    \toprule
    \textbf{Criterion} & \textbf{Cisco CCNx} & \textbf{Fortinet NSE} & \textbf{Fortinet ICS/SCADA} & \textbf{ISA/IEC 62443} \\
    \midrule
    \textbf{Benchmark Category}
      & \textit{IT Networking}
      & \textit{Specialized IT Security}
      & \textit{Conceptual OT Security}
      & \textit{Formal OT Security Standard} \\
    \addlinespace
    \textbf{Knowledge Domain}
      & Cisco Networking
      & Fortinet Security Ecosystem
      & Foundational ICS/OT Concepts
      & The ISA/IEC 62443 Standard \\
    \addlinespace
    \textbf{Question Focus}
      & \textit{Foundational \& General.} IT networking and security principles, with vendor specifics canonical in the industry.
      & \textit{Procedural \& Vendor-specific.} Configuration and management of a proprietary product ecosystem.
      & \textit{Conceptual \& Domain-general.} Core OT security principles and technologies, often independent of a specific vendor.
      & \textit{Standard-specific \& Formal.} Detailed terminology, structure, and prescribed methodologies of the standard. \\
    \bottomrule
  \end{tabularx}
\end{table}

Specifically, questions were collected from freely available content from websites such as \texttt{itexams.com}, \texttt{examtopics.com}, and \texttt{free-braindumps.com}. These platforms were chosen for their extensive and frequently updated question banks, which serve as a proxy for the material covered in the official certification exams. We selected certifications from Cisco, Fortinet, and ISA/IEC as they represent widely recognized industry standards for IT networking and OT systems.

Despite this selection and expert validation of labels, the resulting datasets remain subject to sampling bias. The question pools are drawn from community-driven platforms where users tend to share items they find difficult, surprising, or exam-relevant, rather than a random sample of all possible certification questions or real-world issues. As a consequence, certain topics and question styles may be overrepresented, while scenario-based, free-form, or practical configuration tasks are underrepresented. 

The scraped multiple-choice questions were preprocessed:
\begin{enumerate}
    \item \textbf{De-duplication:} Exact and near-duplicate questions were identified and removed.
    \item \textbf{Formatting Normalization:} Questions with non-standard formats, such as "drag-and-drop" or with reference images, were excluded.
    \item \textbf{Standardization:} All questions, choices, and answers were converted into a unified MMLU-compatible format for seamless integration with evaluation harnesses.
    \item \textbf{Expert Label Verification:} To mitigate label noise arising from the user-generated nature of the source platforms, we manually reviewed all questions and answer keys. Questions with ambiguous wording, inconsistent answers across sources, or suspected wrong solutions were either corrected, based on authoritative documentation, or removed from the dataset entirely. This manual curation step ensures that the final benchmark reflects a high-quality, expert-validated ground truth, although residual noise cannot be fully ruled out.
    \end{enumerate}

\subsection{Baseline Academic Benchmarks}
To compare the results, we include two established cybersecurity benchmarks: 
MMLU Computer Security~\cite{Hendrycks.2021} is a subtask from the widely used Massive Multitask Language Understanding benchmark. It consists of multiple-choice questions designed to test general knowledge in computer security. We use it as a baseline to compare the performance of models on the specialized datasets.

CyberMetric80~\cite{Tihanyi.2024} is a more recent benchmark created using Retrieval-Augmented Generation (RAG) from authoritative sources like NIST standards and research papers. The quality of the content was approved by human experts.

\subsection{Industry Certification Benchmarks}

The core of our evaluation is CyberCertBench, a novel suite of datasets derived from industry-recognized certifications. This approach provides high-quality questions that directly measure the practical knowledge required of professionals across a gradient of specificity. As detailed in Table~\ref{tab:cert_benchmark_analysis}, the benchmarks test distinct types of cybersecurity knowledge. The Fortinet NSE dataset focuses on procedural, vendor-specific configuration, whereas the Fortinet ICS/SCADA dataset assesses broader, conceptual knowledge of the OT security domain. The ISA/IEC 62443 benchmark is the most specialized, testing formal methodology and terminology of that series of standards that address security for \Gls{OT} in automation and control systems. 

\begin{figure}[htp]
  \centering
  \includegraphics[width=\linewidth]{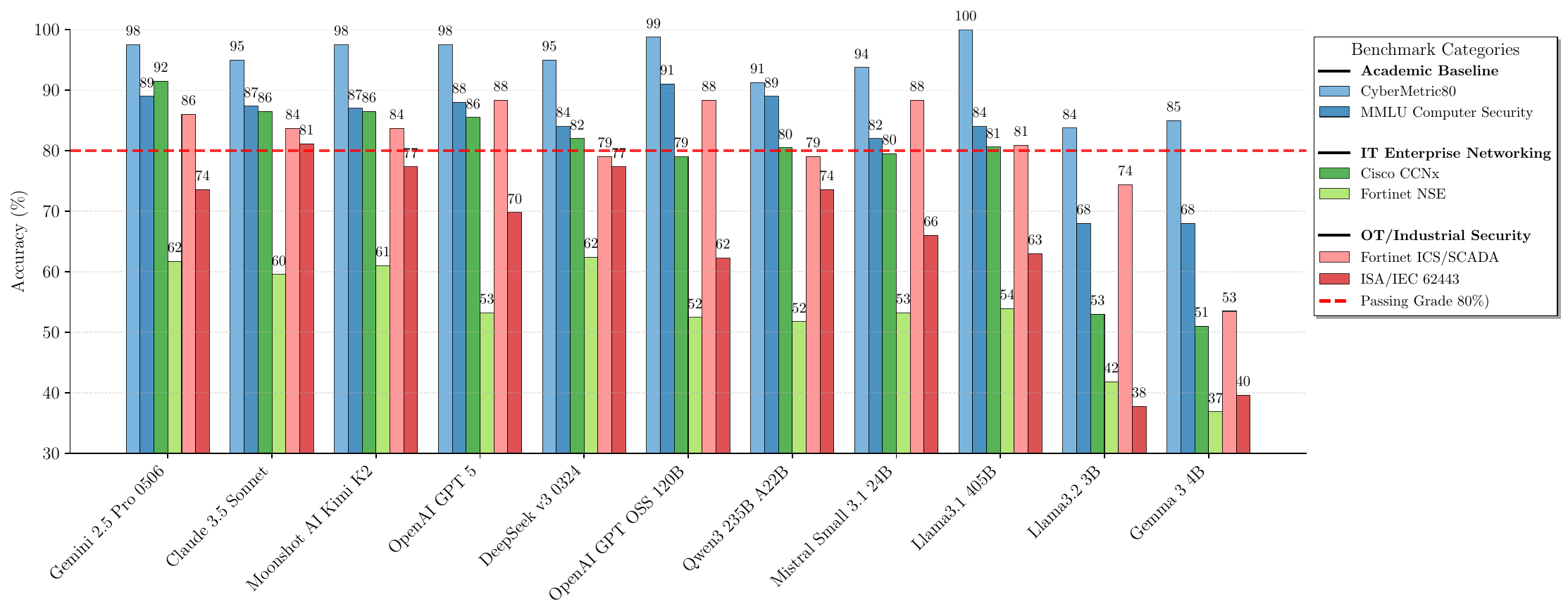}
  \caption{Overview of 5-shot accuracy for a selection of LLMs across six cybersecurity benchmarks ordered by average accuracy. Top proprietary models show  similar performance compared to open-weight competitors, with specialized OT datasets proving more challenging for all models.}
  \label{fig:overall_performance_selected}
\end{figure}

\subsection{Model Selection}
Our model selection aims to cover the state-of-the-art across proprietary and open-weight models of various sizes, leveraging MMLU~\cite{Liang.2023} leaderboards. Selection criteria include a range of models in terms of:
\begin{itemize}
    \item \textbf{Modelsize:} From small to the largest models
    \item \textbf{Region:} Models from the US, Europe and Asia
    \item \textbf{Release Date:} Models released since ChatGPT in 2022
    \item \textbf{Availability:} Proprietary and open-weight models
\end{itemize}

The selection includes the latest proprietary frontier models from leading US-based labs, including OpenAI, Anthropic, and Google. To capture the full performance spectrum, we also include their smaller, efficiency-focused counterparts. The evaluation also features a wide array of open-weight models to analyze the broader ecosystem. This includes foundational model families from Western developers, such as Meta's Llama series, Mistral AI's models, and Google's Gemma family. To provide a global perspective, we have incorporated leading models from Asia, including Alibaba's Qwen series and Moonshot AI's Kimi K2. The recent open-weight contributions from OpenAI (GPT-OSS series) are also included.

\subsection{Experimental Setup and Metrics}
The evaluation is conducted as an MCQA task, where a model is given a question \(q\) and a set of choices \(\mathcal{C}\), from which it must select the correct option(s) \(a\). We use a 5-shot prompting approach, providing five in-domain examples before the actual question to set the context. For all tests, the model temperature is set to zero to ensure deterministic outputs.

To contextualize model performance, we establish a human expert baseline. The passing score for professional certifications like the Cisco exams is typically around 80\%. This score represents the minimum competency expected of a certified professional and serves as a target benchmark.

Performance is measured using exact accuracy. A response is considered correct only if the model identifies all of the correct option(s) and none of the incorrect ones. This strict metric is suitable for the MCQA format where partial credit is not applicable.

\begin{figure}[htp]
  \centering
  \includegraphics[width=\linewidth]{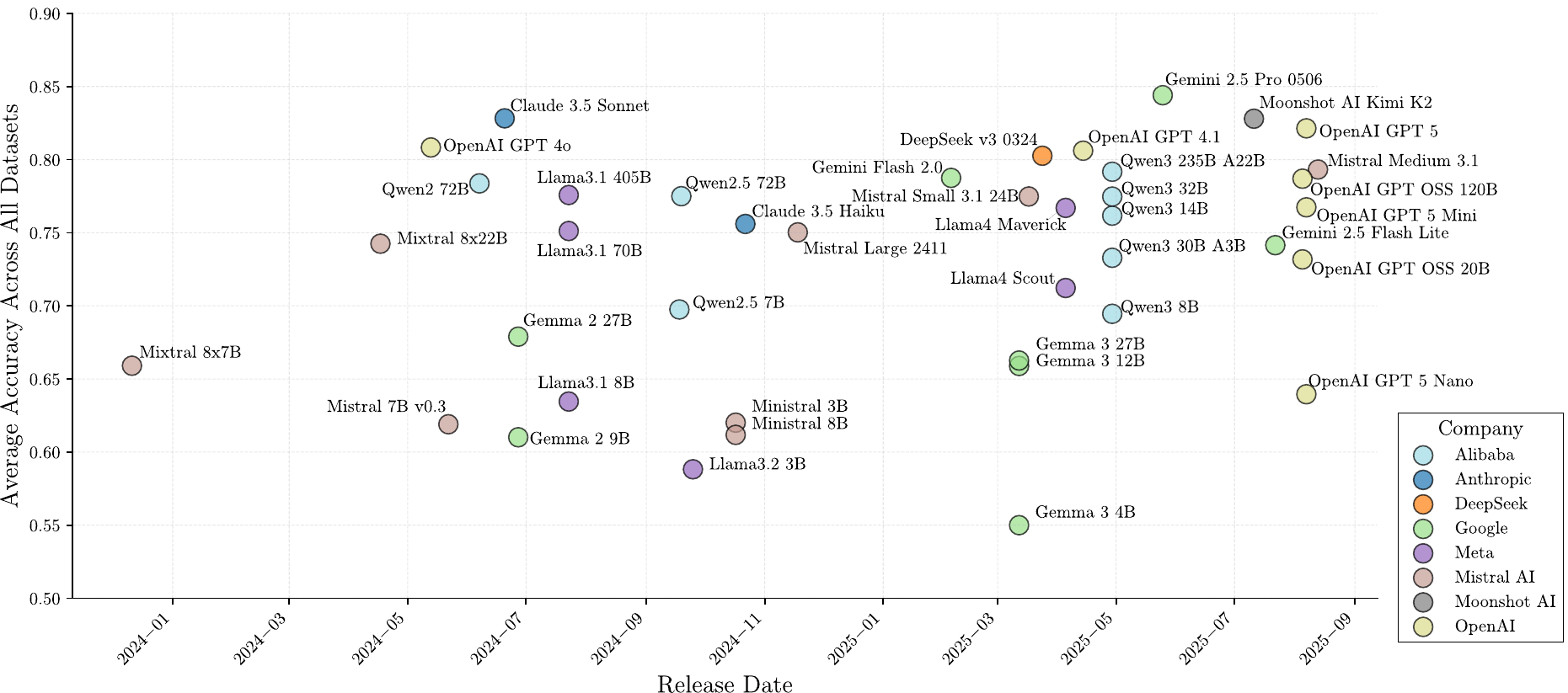}
  \caption{Average accuracy across all certification datasets plotted against release date. The upward trend highlights progress, with smaller, newer models approaching the performance of larger, older models.}
  \label{fig:perf_over_time}
\end{figure}

\subsection{Legal Considerations}
The certification-style questions used in CyberCertBench were obtained via automated scraping from publicly accessible platforms. Our use of this material is limited to internal text-and-data processing for non-commercial scientific research, and based on  access in the sense that no technical protection measures were circumvented. In line with the restrictions on handling IP-sensitive content, we do not publicly redistribute the full question text. The paper reports only aggregate statistics and derived analyses. Access to the underlying questions can be granted to qualified researchers on request.

\section{Quantitative Evaluation}
This section presents a quantitative analysis of LLM performance across cybersecurity benchmarks. First, the overall model accuracies in comparison to human baseline is analyzed. Then the influence of model scale and release date factors on performance is investigated revealing trends in scaling laws and open-weight competition.

\subsection{Performance on Cybersecurity Certifications}
Figure~\ref{fig:overall_performance_selected} presents the performance of a representative set of 11 models across six   cybersecurity benchmarks. The results reveal a performance hierarchy, with large proprietary models such as Gemini 2.5 Pro, Claude 3.5 Sonnet, and OpenAI GPT 5 consistently achieving the highest scores. Notably, top-tier open-weight models, like Moonshot AI's Kimi K2, demonstrate highly competitive performance.

On general IT security benchmarks like CyberMetric80 and MMLU Computer Security, frontier models approach or achieve perfect scores, suggesting these benchmarks are becoming saturated. To contextualize these results, we compare model performance against a human expert baseline, represented by the 80\% passing grade typical for professional certifications. On the IT-focused Cisco certification, top models consistently meet or exceed this professional standard. On the Fortinet ICS/SCADA benchmark, which tests more conceptual OT security knowledge, many top models still achieve a passing grade. This indicates that general concepts in OT security are somewhat accessible. 

However, on the most specialized benchmark, \textbf{ISA/IEC 62443}, which tests formal, standard-specific knowledge, performance collapses. Only a single model, Claude 3.5 Sonnet at 81\%, achieves a passing score. Notably, Claude 4.5 Sonnet has a lower performance on this benchmark and a lower average performance. This shows that while LLMs may have a conceptual understanding at the surface level of OT, they lack the deep, formal knowledge required to be reliable in safety-critical industrial environments.

The challenge is most pronounced on the Fortinet NSE benchmark, which focuses on vendor-specific procedural knowledge. No model comes close to the passing grade, with the highest score being 62.4\%. This indicates that while LLMs trained on web-scale data have acquired a high conceptual IT and OT security knowledge, they lack an understanding of formal standards and vendor-specific operational details. This specialized knowledge remains a significant and measurable frontier.

\subsection{Model Performance Over Time}

Figure~\ref{fig:perf_over_time} analyzes model performance as a function of release date to offer insight into the pace of progress. The results reveal a clear positive overall trend, indicating that the knowledge of LLMs in cybersecurity is advancing. However, this progress is not consistent across all model families, and model sizes.

The Qwen series demonstrates a consistent improvement trajectory from Qwen2 through Qwen3, with newer models maintaining or outperforming older ones. In contrast, the Llama 4 series does not show an improvement over the Llama 3 series on the benchmarks, and the Gemma 3 series exhibits lower average performance than its predecessor, Gemma 2. This suggests that progress is highly dependent on the specific training data used for each model generation, which may not always align with improvements on these specialized tasks. The top of the performance leaderboard also shows Google's Gemini 2.5 Pro (May 2025) surpassing the earlier Claude 3.5 Sonnet (July 2024) as the best-performing model overall for almost one year, while in the open-weight category, DeepSeek v3 and then Kimi K2 surpass Qwen2 72B.

While performance improvements are less profound in the largest models, the increase among smaller, more efficient language models is to be noted. Models in the 14B to 32B parameter range, such as Qwen3 14B/32B and Mistral Small 3.1 24B, show a substantial leap in capability over time, achieving performance levels that are highly competitive with much larger models from previous generations. This trend indicates that progress is driven not by scaling up parameter counts but by advances in training methodology, particularly in data curation. 
\begin{figure}[htp]
  \centering
  \includegraphics[width=\linewidth]{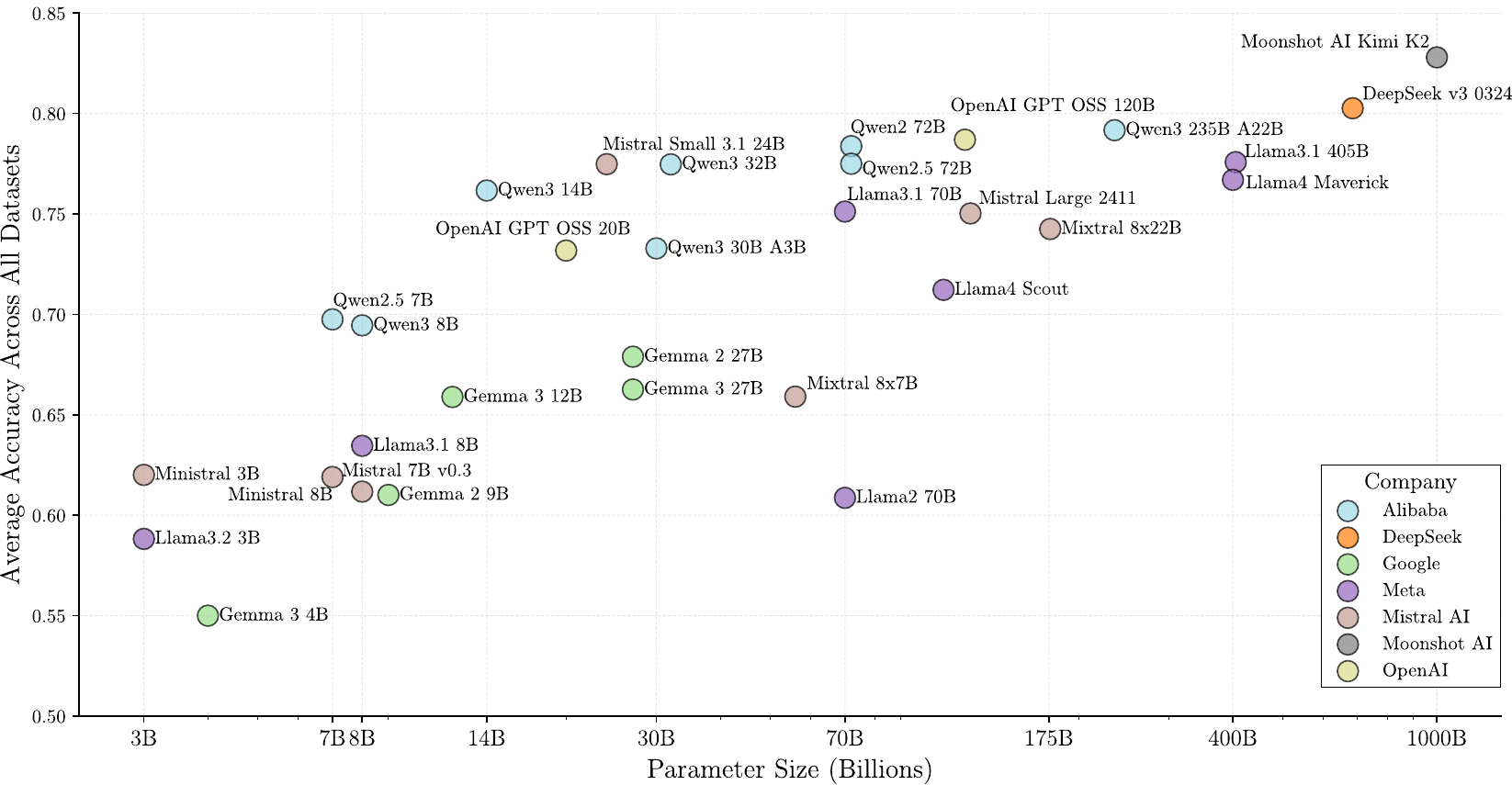}
  \caption{Open-weight model average accuracy across all certification datasets versus parameter size in billions, on a logarithmic scale. The plot shows a clear positive trend where larger models generally perform better, but also highlights significant differences in parameter efficiency among models of similar sizes.}
  \label{fig:accuracy_size}
\end{figure}
\subsection{Scaling Laws: Performance vs. Parameters}

To investigate the relationship between model size and cybersecurity knowledge, the average accuracy of each open-weight model is plotted against its parameter count, as shown in Figure~\ref{fig:accuracy_size}. While this plot illustrates trends in the open-weight ecosystem, the exact parameter counts of the top-performing proprietary models are not public. The plot reveals a strong positive correlation between model size and average performance. The largest open-weight models, such as Moonshot AI's Kimi K2 (~1 trillion parameters) and DeepSeek v3 0324, occupy the top-right quadrant.

However, the distribution also highlights significant variations in \textit{parameter efficiency}. Models of similar size exhibit a wide performance spread. A clear example is the Llama2 70B model, which is substantially outperformed by more recent models in the same size class, such as Qwen2 72B and Llama3.1 70B, demonstrating generational progress.

This trend is even more pronounced when comparing models across different size classes. For instance, models such as Qwen3 14B and Mistral Small 3.1 24B achieve average accuracies that are highly competitive with much larger models like Llama 3.1 405B. This demonstrates that improvements in parameter efficiency allow smaller models to achieve performance on par with counterparts an order of magnitude larger in these specialized domains.

\subsection{Evaluating on a More Challenging "PRO" Benchmark}
\begin{figure}[htbp]
\CenterFloatBoxes     
\begin{floatrow}

\killfloatstyle          
\ttabbox%
  {%
    \begin{tabular}{@{}lrrr@{}}
      \toprule
      \textbf{Dataset} & \textbf{\#Tot} & \textbf{\#Easy} & \textbf{\#PRO} \\
      \midrule
      CCNx                & 200 & 109 &  91 \\
      Fort. ICS/SCADA  &  43 &  29 &  14 \\
      ISA/IEC 62443       &  53 &  24 &  29 \\
      Fortinet NSE        & 141 &  55 &  86 \\
      \bottomrule
    \end{tabular}
  }%
  {%
    \caption{Statistics of the PRO benchmark showing the original number of
    questions, easy questions removed, and remaining in the PRO version.}%
    \label{tab:pro_stats}%
  }

\ffigbox%
  {\includegraphics[width=0.4\textwidth]{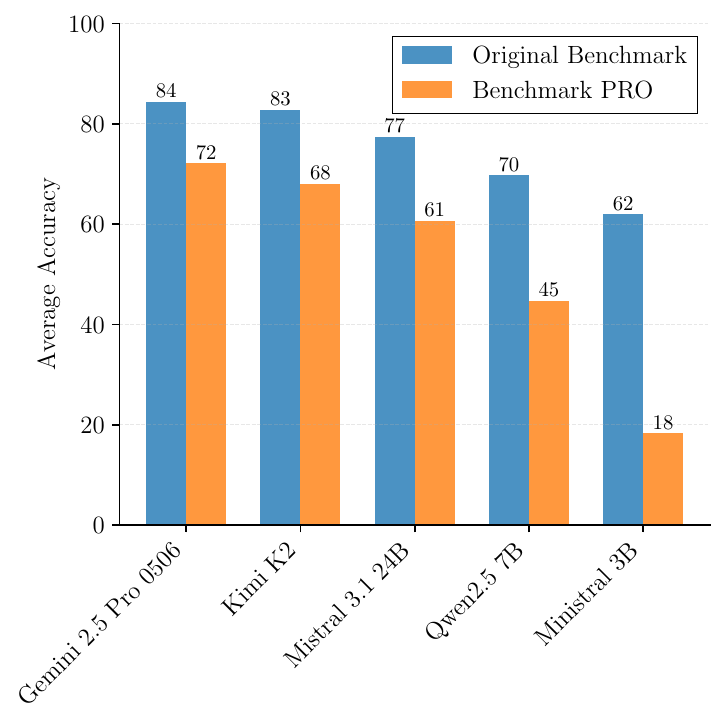}}%
  {%
    \caption{Average model accuracy on the original benchmarks versus the
    ``PRO'' version on selected models.}%
    \label{fig:pro_vs_original}%
  }

\end{floatrow}
\end{figure}

As state-of-the-art models approach saturation on standard benchmarks, their utility for differentiating between top performers is reduced. Inspired by the MMLU-Pro~\cite{Wang.2024b}, we created a more difficult "PRO" version of the benchmark suite through a data-driven curation process that filters out easy questions. A question is classified as easy if it is answered correctly by at least two out of three small models: Ministral 3B, Gemma 3 4B, and Llama3.2 3B. As shown in Table~\ref{tab:pro_stats}, this filtering process removed approximately 50\% of the questions from the certification-based benchmarks. The removal rate varied from 39.0\% for the Fortinet NSE dataset to 67.4\% for the Fortinet ICS/SCADA dataset, indicating differences in baseline difficulty across the domains.

While performance drops universally, the magnitude of the drop is not uniform and correlates with model capability. Top-tier models like Gemini 2.5 Pro and Moonshot AI Kimi K2 show a low performance decrease. In contrast, smaller models experience a higher performance degradation. The accuracy of Qwen2.5 7B falls by 25 percentage points, while the small baseline model, Ministral 3B, falls by 44 points from 62\% to just 18\%, indicating that it still answers some questions correctly that the other two small models cannot. 

\section{Interpretable Difficulty Analysis}
\begin{figure}[h]
    \centering
    \includegraphics[width=1\linewidth]{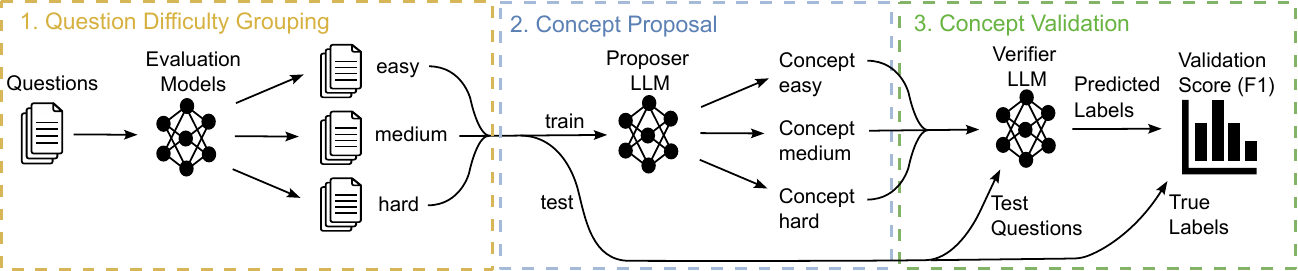}
    \caption{The Proposer-Verifier framework workflow. The process involves (1) grouping questions by difficulty, (2) using a Proposer LLM to generate difficulty descriptions from a training set, and (3) using a Verifier LLM to validate these descriptions by classifying questions from a held-out test set to produce the final validation statistics.}
    \label{fig:proposer-verifier-flowchart}
\end{figure}

The quantitative analysis reveals a significant discrepancy between model performance on IT and OT benchmarks. However, these aggregate accuracy scores do not explain the \emph{reasons} why certain questions or entire domains are more difficult for LLMs. To understand these underlying factors, we conducted a qualitative analysis of the datasets. Inspired by recent work that uses LLMs to describe dataset differences~\cite{Zhong.2022b}, we implement a Proposer-Verifier framework. To the best of the authors' knowledge,  this is the first method that adapts Item Response Theory (IRT) by first identifying groups of questions with varying difficulty~\cite{Wu.2024a} and then using LLMs to generate and validate interpretable descriptions of these difficulty groups.

The methodology consists of three main steps:
\begin{enumerate}
    \item \textbf{Difficulty Grouping:} Rank all questions based on their empirical difficulty as average accuracy across all models. Partition them into a set number of distinct difficulty groups, like hard, medium, and easy.
    \item \textbf{Concept Proposal:} A LLM (the "Proposer") takes questions from the \textit{training split} of each difficulty group and generates a natural language description of the distinguishing characteristics of each group.
    \item \textbf{Concept Validation:} To validate the quality and utility of these descriptions, we use them as a classification rubric. A LLM (the "Verifier") is tasked with predicting the difficulty group of questions from a held-out \textit{test split}, based solely on the generated descriptions.
\end{enumerate}
This Proposer-Verifier method, as shown in Figure~\ref{fig:proposer-verifier-flowchart}, bridges quantitative difficulty signals with qualitative explanations, yielding natural language explanations into model performance. The following section provides a more detailed description of the method.

\subsection{Mathematical Formulation}
Let \( \mathcal{Q} = \{q_1, \dots, q_N\}\) be the set of questions in a benchmark, where $N$ is the total number of questions. Let \(\mathcal{M}=\{m_1, \dots, m_{M}\}\) be the set of evaluated LLMs. For each question \(q_i\) and model \(m_j\), we define the observed binary response
\[
r_{ij} =
\begin{cases}
1 & \text{if model } m_j \text{ answers } q_i \text{ correctly},\\
0 & \text{otherwise.}
\end{cases}
\]

\paragraph{\textbf{Question Difficulty Grouping}}
The difficulty \(p_i\)  of a question \(q_i\) is its average accuracy across the set of models:
\[
p_i = \frac{1}{|\mathcal{M}|}\sum_{j=1}^{|\mathcal{M}|} r_{ij} \in [0,1].
\]
A smaller \(p_i\) number indicates a harder question. We rank all questions in \(\mathcal{Q}\) based on their \(p_i\) values and partition the sorted list into \(G\) equal-sized quantile groups, \(\mathcal{Q}_1, \mathcal{Q}_2, \dots, \mathcal{Q}_G\), where \(\mathcal{Q}_1\) contains the \(\lfloor N/G \rfloor\) hardest questions (lowest \(p_i\)) and \(\mathcal{Q}_G\) contains the \(\lceil N/G \rceil\) easiest questions (highest \(p_i\)). For the analysis, we use \(G=3\) to create \textit{hard}, \textit{medium}, and \textit{easy} groups.

\begin{table}[htbp]
\caption{Mean and standard deviation of F1 score across the different analysis methods (\%). Best values are bold.}
\label{tab:f1_comparison_comprehensive}
\begin{tabularx}{\textwidth}{l *{6}{Y}}

\toprule
Method & ISA/IEC 62443 & Fortinet NSE & Fort. ICS/SCADA & CCNx & MMLU Sec & Cyber- Metric80 \\
\midrule
Random Guess & 33.3 & 33.3 & 33.3 & 33.3 & \textbf{33.3} & \textbf{33.3} \\
Baseline     & 44.4±7.2 & 39.6±0.6 & 44.1±14.2 & 36.5±2.3 & 28.0±10.4 & 27.2±0.4 \\
Difficulty   & 45.9±10.3  & 39.8±0.7 & 42.0±10.1 & \textbf{43.9±6.8} & 28.4±4.3 & 23.0±4.7 \\
Synthesized  & \textbf{47.7±11.8} & \textbf{41.9±1.5} & \textbf{48.9±13.2} & 40.2±0.9 & 31.5±4.4 & 23.9±9.2\\
\bottomrule
\end{tabularx}
\end{table}

\paragraph{\textbf{Proposer-Verifier Framework and Baselines}}  
To validate the analysis, each difficulty group \(\mathcal{Q}_g\) is randomly partitioned into a training set \(\mathcal{Q}_g^{\text{train}}\) and a held-out test set \(\mathcal{Q}_g^{\text{test}}\) with for example an often used 80/20 split. We then generate difficulty descriptions using different "Proposer" methods:

\begin{enumerate}
    \item \textbf{Baseline Proposer (Question-Only):} As a baseline, an LLM \(\Phi_B\) generates descriptions \(\mathcal{D}_B\) by analyzing only the question text from the training sets, without being provided the pre-calculated difficulty labels. This tests the ability to infer difficulty from content alone.
    \[
    \Phi_B(\{\text{sample}(S) \mid S \in \{\mathcal{Q}_1^{\text{train}}, \dots, \mathcal{Q}_G^{\text{train}}\}\}) \to \mathcal{D}_B
    \]

    \item \textbf{Analysis Proposer (Single-Run):} The primary analysis method uses an LLM \(\Phi_A\) that takes samples from the \textit{labeled} difficulty groups in the training sets to generate a set of natural language descriptions \(\mathcal{D}_A\).
    \[
    \Phi_A(\{\text{sample}(S) \mid S \in \{\mathcal{Q}_1^{\text{train}}, \dots, \mathcal{Q}_G^{\text{train}}\}\}) \to \mathcal{D}_A
    \]
    This is repeated over \(k\) independent runs to account for model variability, producing \(k\) sets of descriptions \(\{\mathcal{D}_A^{(1)}, \dots, \mathcal{D}_A^{(k)}\}\).

    \item \textbf{Synthesizer Proposer (Multi-Run):} To create a more robust set of descriptions, a third LLM mapping \(\Phi_S\) acts as a synthesizer. It takes the \(k\) sets of descriptions from the Analysis Proposer as input and generates a single, unified set of descriptions \(\mathcal{D}_S\). This can also be repeated several times.
    \[
    \Phi_S(\{\mathcal{D}_A^{(1)}, \dots, \mathcal{D}_A^{(k)}\}) \to \mathcal{D}_S
    \]
    For the following experiments, we use \(k=3\) runs for synthesis.
\end{enumerate}

\paragraph{\textbf{Verifier (Description Validation)}}
The Verifier is an LLM-based classifier \(\Phi_V\) that uses a generated set of descriptions \(\mathcal{D}\) to predict the difficulty group of an unseen question \(q\) from the test set.
\[
\Phi_V(q, \mathcal{D}) \to g' \in \{1, \dots, G\} \quad \text{where } q \in \bigcup_{g=1}^G \mathcal{Q}_g^{\text{test}}
\]
The effectiveness of each Proposer method is quantified by the classification performance (F1-score) of the Verifier when using its corresponding descriptions (\(\mathcal{D}_B\), an averaged over \(\mathcal{D}_A^{(i)}\), and \(\mathcal{D}_S\)) on the held-out test sets. For the Proposer, Verifier, and Synthesizer tasks, we use OpenAI's GPT-5.

\subsection{Results and Analysis}

\begin{figure}[htbp]
  \centering
  \includegraphics[width=\linewidth]{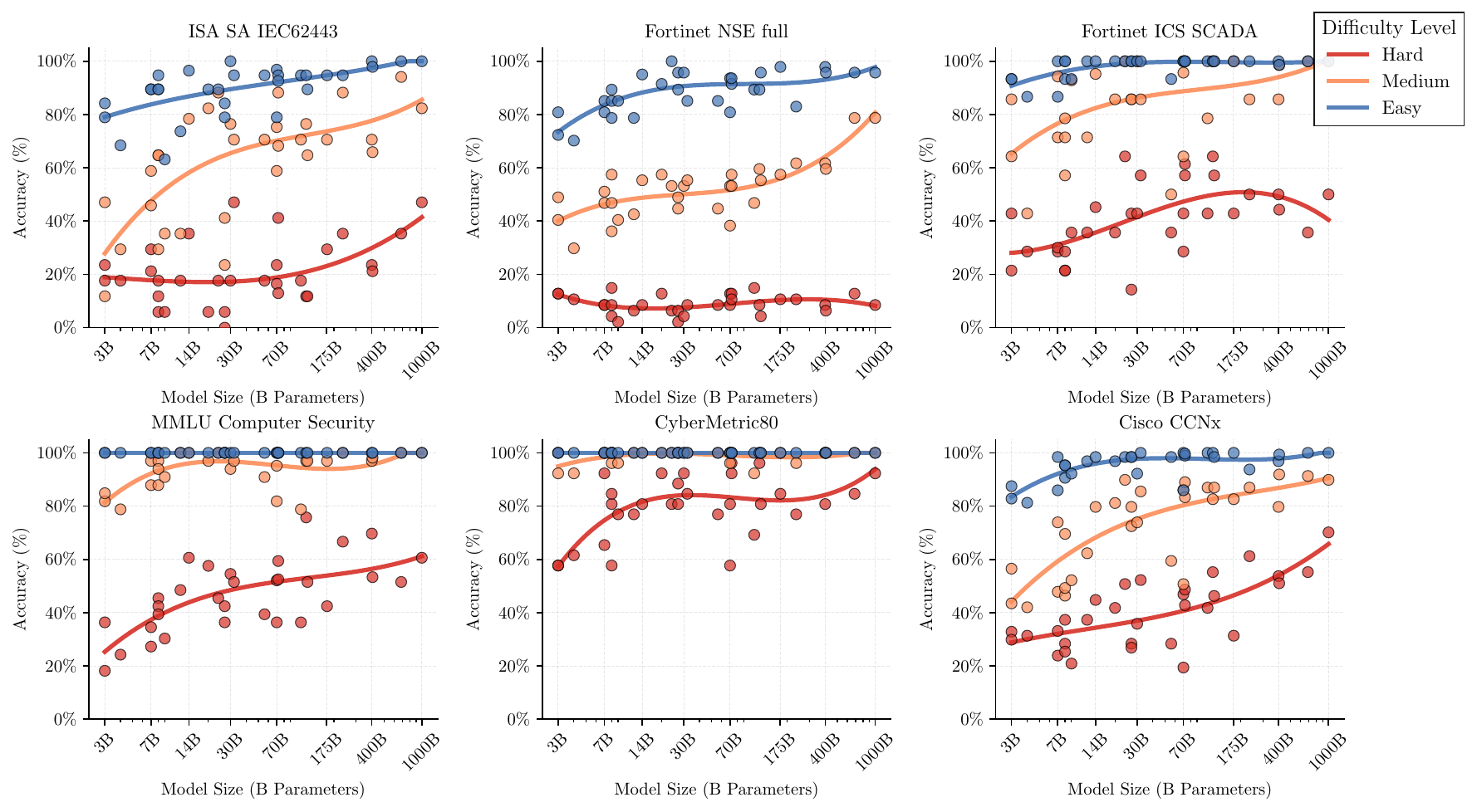}
  \caption{Question set accuracy as a function of model size (log scale) across three difficulty tiers (Hard, Medium, Easy) for each benchmark with a regression fitted to show the scaling trend. The top row shows the OT security benchmarks, the bottom row shows the IT-focused benchmarks.}
  \label{fig:difficulty_scaling_plot}
\end{figure}

This section presents the qualitative analysis based on the Proposer-Verifier framework. We first examine how capabilities emerge as model scale increases across different difficulty tiers within the benchmarks. Second, we assess how well the generated difficulty descriptions can be used to classify unseen questions.

\paragraph{Emergent Knowledge Capabilities and Difficulty Scaling}
To study how cybersecurity knowledge scales with model size, we partitioned the questions in each benchmark into three tiers, hard, medium, and easy, based on average accuracy across models. Figure~\ref{fig:difficulty_scaling_plot} shows accuracy versus parameter count for these tiers. Performance generally improves with scale, but the scaling behavior and gaps between tiers vary by benchmark.

For the three OT-focused benchmarks (top row), the performance gap between easy, medium, and hard questions is wide and persistent. On ISA/IEC 62443 and Fortinet NSE, even the largest models stay below 40\% accuracy on the hardest questions. The nearly flat curve on the hardest ISA/IEC 62443 items suggests that increasing model size alone does not add the formal, standard-specific knowledge required; specialized OT knowledge remains a frontier.

The general IT security benchmarks (bottom row) show different behavior. MMLU Computer Security and CyberMetric80 exhibit benchmark saturation: accuracy on easy and medium questions clusters near 100\%, even for smaller models, so difficulty levels are poorly separated and most items are too simple to differentiate modern LLMs. The Cisco CCNx benchmark lies between these extremes, with clearer separation between all three tiers and consistent gains with model scale, indicating a more informative and learnable domain.

\paragraph{Validating the Qualitative Analysis Methodology}
To evaluate the usefulness of the generated difficulty descriptions, we used them as a rubric for classifying a held-out set of questions. Table~\ref{tab:f1_comparison_comprehensive} reports F1-scores for a Verifier LLM using three description sets: a baseline derived from question text alone, descriptions from single analysis runs, and descriptions synthesized from multiple runs.

On average, all methods outperform the random baseline of 33.3\%, indicating that the descriptions capture signals of difficulty, but only when the benchmarks have sufficient separation between difficulty tiers (as in Figure~\ref{fig:difficulty_scaling_plot}). For MMLU Computer Security and CyberMetric80, this separation is weak, and clustering by difficulty fails. The baseline “Questions Only” descriptions reach an average F1-score of 36.7\%, the single-run analysis 37.2\%, and the synthesized descriptions 39.0\%. Large standard deviations mean differences are not always statistically clear, but the trend suggests that multi-run synthesis yields the most robust and general descriptions.

A notable outlier is CyberMetric80, where all methods achieve F1-scores below the random baseline. This benchmark contains too few genuinely hard questions to support meaningful difficulty tiers. Overall, the validation indicates that the framework produces interpretable descriptions that reflect difficulty patterns, but only for benchmarks with a sufficiently broad and challenging item distribution.

\paragraph{Natural Language Descriptions}
The synthesized difficulty descriptions reveal a progression of required capabilities. Easy questions primarily test factual recall of foundational, single-concept definitions such as standard acronyms or default port numbers. Medium questions emphasize application, requiring models to map principles to scenarios and discriminate between plausible options. The hardest questions demand specialization, including vendor-specific command semantics, formal standards, and operational nuances.

These findings show that current LLMs encode a large body of general IT security facts, but their limitations emerge on specialized, context-dependent knowledge that underpins professional practice. Full natural language descriptions for each benchmark are provided in Appendix~\ref{appendix:difficulty_descriptions}.

\section{Discussion and Limitations}
Our findings empirically demonstrate that LLM knowledge in cybersecurity follows a gradient of specificity. Leading models exhibit expert-level knowledge on canonical IT certifications like Cisco CCNx, but their performance degrades on vendor-specific IT benchmarks such as Fortinet NSE and collapses on tasks requiring formal, standards-based OT knowledge like ISA/IEC 62443. This gradient likely reflects the composition of web-scale pre-training corpora, which contain abundant general IT concepts but relatively little proprietary documentation and formal standards~\cite{10634156}. As a result, strong performance on general benchmarks is not a reliable indicator of capability in specialized, context-dependent tasks, especially in critical domains.

Beyond the IT/OT knowledge gap, our analysis highlights progress in model efficiency. While performance gains for very large models appear to be saturating, newer models in the 14B–32B parameter range rival the performance of previous-generation models an order of magnitude larger, suggesting ongoing improvements in training methodology and data curation. This increase in parameter efficiency has implications for energy-efficient AI and broader access to capable models.

Methodologically, this work introduces a Proposer-Verifier framework that generates and validates interpretable, natural-language descriptions of question difficulty. This provides a way to move beyond aggregate scores and investigate the causes of model successes and failures.

A key limitation stems from our data sources for the certification benchmarks. The questions were curated from publicly accessible, community-driven websites. Although these platforms approximate official exam content, their reliance on user-submitted questions and answers introduces a risk of inaccuracies and outdated material. To mitigate this risk, we performed a manual expert review of all questions and answer keys. Items with ambiguous wording, conflicting answers across sources, or insufficiently justified solutions were corrected based on authoritative documentation, or removed. After this expert curation, some residual label noise and selection bias may remain. 

The Proposer-Verifier framework is also limited by its self-referential nature: we use a frontier LLM to analyze the capabilities of other LLMs. The Proposer may share the same knowledge gaps as the evaluated models, potentially misattributing the source of difficulty. The quality of the analysis is therefore tied to the specific model used for the Proposer task. In addition, the non-deterministic behavior of language models and their sensitivity to prompts can lead to variation between runs, introducing qualitative uncertainty compared to standard quantitative metrics.

Future work should address these limitations by improving the robustness and stability of the qualitative analysis, for example through systematic hyperparameter tuning and more constrained prompting. Methodological extensions such as iterative refinement of descriptions or soft prompting could further reduce variance and sensitivity to prompt design.

\section{Conclusion}
The present paper introduces CyberCertBench, a new suite of benchmarks derived from professional certifications, to evaluate LLM knowledge across a gradient of specificity from general IT to specialized OT security. Our central finding is that LLM performance degrades significantly as the required knowledge shifts from web-prevalent concepts to the vendor-specific procedures and formal standards that define professional expertise. While leading models achieve passing grades on general IT certifications, they consistently fail on benchmarks testing deep, specialized knowledge, particularly in vendor-specific procedures and formal OT standards, posing a risk for their use in critical infrastructure environments.

Our analysis of scaling and release date trends further reveals that while larger models perform generally better, performance increases are diminishing. Smaller LLMs, on the other hand, have shown a large performance increase in the last year. Therefore, the knowledge capabilities are becoming more parameter-efficient. Methodologically, this work contributes both the publicly available OT benchmarks themselves and a novel Proposer-Verifier framework. The analysis reveals a capability progression, from proficiency in foundational factual recall and applied reasoning to a significant performance drop when faced with the specialized demands of vendor-specific or industrial standards-based knowledge.
\newpage

\bibliographystyle{unsrtnat}
\bibliography{bibliography}
\clearpage

\appendix

\section*{Appendices}

\section{Synthesized Descriptions of Question Difficulty}
\label{appendix:difficulty_descriptions}

The following tables provides the short and complete, synthesized natural language descriptions of the characteristics of easy, medium, and hard questions for each benchmark, as generated by the qualitative analysis framework.

\begin{table}[htp]
  \caption{Primary Drivers of Question Difficulty Based on Qualitative Analysis.}
  \label{tab:difficulty_drivers}
\begin{tabularx}{\textwidth}{p{3cm} Y Y }
    \toprule
    \textbf{Difficulty Driver} & \textbf{Synthesized Description} & \textbf{Example Benchmark} \\
    \midrule
    \textbf{Factual Recall} \newline \textit{(Low Difficulty)} 
    & Tests recognition of single-concept definitions, standard terms, and widely known facts. Success relies on direct knowledge retrieval with minimal reasoning.
    & \textbf{MMLU Security:} Identifying a web proxy's function. \newline \textbf{CyberMetric80:} Naming the purpose of a WAF. \\
    \addlinespace
    \textbf{Applied Reasoning} \newline \textit{(Medium Difficulty)} 
    & Requires mapping concepts to scenarios, often involving light troubleshooting, multi-select formats, or discriminating between plausible options.
    & \textbf{CCNx:} Interpreting interface counters or compressing an IPv6 address. \\
    \addlinespace
    \textbf{Vendor-Specific Semantics} \newline \textit{(High Difficulty)} 
    & Demands precise knowledge of a specific vendor's product line, CLI syntax, proprietary features, and nuanced operational behaviors.
    & \textbf{Fortinet NSE:} Knowing SD-WAN policy route precedence rules. \\
    \addlinespace
    \textbf{Standard-Specific Formalism} \newline \textit{(High Difficulty)} 
    & Requires familiarity with the specific terminology, structure, and process phases defined within a formal standard like ISA/IEC 62443.
    & \textbf{ISA/IEC 62443:} Identifying which activities belong to the "assess" phase of the IACS security lifecycle. \\
    \bottomrule
  \end{tabularx}
\end{table}

\begin{longtable}{p{3cm} p{12.5cm}}
  \caption{Synthesized Difficulty Characteristics for Each Benchmark.} \\
  \toprule
  \textbf{Benchmark} & \textbf{Synthesized Difficulty Characteristics} \\
  \midrule
\endfirsthead
\multicolumn{2}{c}%
{{\bfseries \tablename\ \thetable{} -- continued from previous page}} \\
  \toprule
  \textbf{Benchmark} & \textbf{Synthesized Difficulty Characteristics} \\
  \midrule
\endhead
  \bottomrule
\endfoot
  \textbf{Cisco CCNx} & 
  \textbf{Easy:} Short, definition-focused items with one clear answer covering fundamentals like RFC1918, VLAN broadcast domains, AAA/firewalls, and high-level TCP/UDP; minimal vendor syntax or calculations. \newline
  \textbf{Medium:} Conceptual application with light troubleshooting: interpret simple counters, pick non-overlapping channels, compress IPv6, understand flooding scope, and basic CDP/DTP/PoE nuances. \newline
  \textbf{Hard:} Multi-constraint, multi-step synthesis requiring exact Cisco defaults and CLI semantics: STP/RSTP states, PortFast effects, DTP/LACP mode interplay, constrained VLSM, and detailed counter-based troubleshooting. \\
  \addlinespace
  \textbf{MMLU Security} &
  \textbf{Easy:} Short, single-concept, fact-based items on definitions, tools (e.g., Nmap, Tor), CIA triad, OSI/wireless basics, and buffer overflow fundamentals; minimal reasoning with clear distractors. \newline
  \textbf{Medium:} Scenario-oriented questions requiring light application and discrimination: forward secrecy, PKI/HTTPS enhancements (DV/EV, stapling, pinning), IPSec vs TLS, and fuzzing styles. \newline
  \textbf{Hard:} Multi-step, cross-domain reasoning with formal crypto and system/protocol nuances: TLS/Kerberos semantics, Merkle–Damgård length extension, PRF/CPA subtleties, and system architecture specifics (e.g., NaCl validator rules, OKWS FD passing). \\
  \addlinespace
  \textbf{Fortinet NSE} & 
  \textbf{Easy:} Single-concept, factual recall or basic diagnostics with clear cues, minimal ambiguity, and little to no cross-feature dependency; core networking/security fundamentals and straightforward Fortinet basics. \newline
  \textbf{Medium:} Two to three related concepts with moderate scenarios and multi-selects; mapping definitions to product behavior and workflows across SD-WAN, AV modes, FMG ADOMs, SSL VPN, and FSSO/LDAP. \newline
  \textbf{Hard:} Scenario-heavy, multi-select items requiring precise vendor defaults, precedence rules, and internals across multiple Fortinet products and cloud; integrated troubleshooting (FAZ indexing, FMG HA, SIP ALG vs helper, ZTNA). \\
  \addlinespace
  \textbf{Fortinet ICS/SCADA} & 
  \textbf{Easy:} Single-fact, foundational recall with clear, unambiguous stems and canonical answers (CIA, TCP flags, MTU, basic tool facts) and simple distractors. \newline
  \textbf{Medium:} Applied recall and conceptual mapping (IPsec modes, default-deny, ICMP, Modbus, tool roles) with moderate ambiguity (negative stems, plausible distractors) but grounded in standard material. \newline
  \textbf{Hard:} Specialized ICS/SCADA standards and vendor/OS minutiae (IEC 62443, CVSS taxonomy, AH/ESP fields, Windows/nmap defaults) with higher ambiguity (close counts, synonyms, unit mismatches) requiring synthesis and elimination. \\
  \addlinespace
  \textbf{CyberMetric80} & 
  \textbf{Easy:} Direct recall of basic definitions and purposes with clear single-correct answers and low-plausibility distractors; core networking/security and governance terms (e.g., ARP, TLS, MFA, WAF). \newline
  \textbf{Medium:} Conceptual understanding and light application; select the primary purpose/control/role among plausible options across PKI, networking ports/handshakes, governance/risk/BCP/PCI, and common attacks. \newline
  \textbf{Hard:} Multi-step application with precise syntax and evolving-standards nuance; includes CIDR/IPv6/XOR, exact tool flags (nmap -sS), specialized crypto/RNG (SRTP AES-GCM, DRBG state protection, DLIES), and policy subtleties. \\
  \addlinespace
  \textbf{ISA/IEC 62443} & 
  \textbf{Easy:} Foundational networking and security/ICS facts (e.g., OSI, default deny, phishing, management risk role) with clear answers and obvious distractors; primarily factual recall. \newline
  \textbf{Medium:} Recall plus light application in ICS contexts (62443 family, TRs like patching, IDS limits, crypto basics, patching priority). Includes best-practice judgments. \newline
  \textbf{Hard:} Precise, standards-heavy ISA/IEC 62443 topics (zones/conduits, lifecycle assess phase, access control variables), sector-specific standards (API 1164), ISASecure governance, and multi-select with qualifiers and close distractors. \\
\end{longtable}

\section{Prompts for Qualitative Difficulty Analysis}
\label{appendix:prompts}

This appendix contains the full prompts used in the Proposer-Verifier framework for the qualitative analysis of question difficulty.

\begin{tcolorbox}[
  breakable,
  colback=white,
  colframe=black,
  boxrule=1pt,
  sharp corners,
  title=\textbf{Prompt for Baseline},
  coltitle=white,
  colbacktitle=black
]
\begin{lstlisting}[breaklines]
You are an expert in cybersecurity education and assessment. You have been given a large set of multiple choice
questions from the {exam_name} dataset. Your task is to analyze these questions and create difficulty level 
descriptions that could be used to classify questions into {num_groups} difficulty levels.

TRAINING QUESTIONS:
{all_questions}

Based on your analysis of these questions, create descriptions for {num_groups} difficulty levels that capture
the key characteristics that would distinguish questions of different complexity. Consider factors like:

1. **Knowledge Types**: What types of knowledge do different questions test?
2. **Complexity Factors**: What makes some questions more challenging than others?
3. **Content Areas**: Are there topic patterns that correlate with difficulty?
4. **Question Structure**: How do questions differ in construction and presentation?

At the end, provide a JSON summary with the following structure:

{{
  "baseline_analysis": {{
    "easiest_group_characteristics": "Clear description of what would define the easiest questions",
    "medium_group_characteristics": "Clear description of what would define medium difficulty questions", 
    "hardest_group_characteristics": "Clear description of what would define the hardest questions",
    "difficulty_progression": "How questions would change from easy to medium to hard",
    "key_differences": ["difference 1", "difference 2", "difference 3"],
    "methodology": "Brief explanation of how you determined these difficulty levels"
  }}
}}
\end{lstlisting}
\end{tcolorbox}

\begin{tcolorbox}[
  breakable,
  colback=white,
  colframe=black,
  boxrule=1pt,
  sharp corners,
  title=\textbf{Prompt for Single-Run Difficulty Analysis (Proposer)},
  coltitle=white,
  colbacktitle=black
]
\begin{lstlisting}[breaklines]
You are an expert in cybersecurity education and assessment. 
Analyze the following sets of multiple choice questions from the {exam_name} dataset, grouped by difficulty level:

DIFFICULTY GROUP OVERVIEW:
{groups_summary}

HARDEST GROUP ({hardest_group}):
{analysis_results[hardest_group]['formatted_questions']}
##############
MEDIUM GROUP ({medium_group}):
{analysis_results[medium_group]['formatted_questions']}
##############
EASIEST GROUP ({easiest_group}):
{analysis_results[easiest_group]['formatted_questions']}

Please analyze the key differences across all three difficulty groups and provide insights into:

1. **Knowledge Types**: What types of knowledge do the question groups test?
2. **Complexity Factors**: What makes questions challenging across difficulty levels?
3. **Content Areas**: Are there specific cybersecurity domains or topics that appear more in one group?
4. **Question Structure**: How do the questions differ in their construction?
5. **Difficulty Progression**: What patterns emerge as questions progress from easy to hard?

At the end, provide a JSON summary with the following structure:
{
  "analysis_summary": {
    "easiest_group_characteristics": "...",
    "medium_group_characteristics": "...",
    "hardest_group_characteristics": "...",
    "difficulty_progression": "...",
    "key_differences": ["...", "...", "..."],
    "knowledge_gaps": "...",
    "recommendations": "..."
  }
}
\end{lstlisting}
\end{tcolorbox}

\begin{tcolorbox}[
  breakable,
  colback=white,
  colframe=black,
  boxrule=1pt,
  sharp corners,
  title=\textbf{Prompt for Multi-Run Synthesis (Synthesizer)},
  coltitle=white,
  colbacktitle=black
]
\begin{lstlisting}[breaklines]
You are tasked with combining multiple responses into a single, cohesive response.
Below, I will provide several responses from different analysis runs.
Your goal is to identify common themes, reconcile differences, and combine the information
into a unified response.
Be sure to preserve all key insights from each trace and ensure the final output is logically
consistent and comprehensive.

{rollouts}

Output Format:
Combine all the provided responses into a new, comprehensive, complete, and unified response, prefixed by
"# UNIFIED RESPONSE".
Your response should not be much longer than the original responses.

At the end, provide a JSON summary with the following structure:

{{
  "analysis_summary": {{
    "easiest_group_characteristics": "Clear unified description of what defines the easiest questions",
    "medium_group_characteristics": "Clear unified description of what defines the medium difficulty questions", 
    "hardest_group_characteristics": "Clear unified description of what defines the hardest questions",
    "difficulty_progression": "How questions change from easy to medium to hard",
    "key_differences": ["difference 1", "difference 2", "difference 3"],
    "knowledge_gaps": "Main knowledge gaps revealed across difficulty levels",
    "recommendations": "Suggestions for improving LLM performance across difficulty levels"
  }}
}}
\end{lstlisting}
\end{tcolorbox}

\begin{tcolorbox}[
  breakable,
  colback=white,
  colframe=black,
  boxrule=1pt,
  sharp corners,
  title=\textbf{Prompt for Difficulty Classification (Verifier)},
  coltitle=white,
  colbacktitle=black
]
\begin{lstlisting}[breaklines]
You are an expert in question classification. Given the following three descriptions and a question,
predict which class the question belongs to.

CLASS  DESCRIPTIONS:

CLASS 1:
{easiest_description}

CLASS 2:
{medium_description}

CLASS 3:
{hardest_description}

QUESTION TO CLASSIFY:
{question_text}

Based on the characteristics described above, which class does this question belong to? 
After careful consideration, respond with only one of: "CLASS 1", "CLASS 2", or "CLASS 3"
\end{lstlisting}
\end{tcolorbox}
\newpage
\section{Full Benchmark Accuracy}
\label{app:accuracy}

\begin{longtable}{lcccccc}
\caption{Mean accuracy (\%) for every benchmark and model in alphabetical order.}
\label{tab:model_benchmark_accuracy}\\
\toprule
& \makecell{CyMetric\\80}
& \makecell{MMLU\\CompSec.}
& \makecell{Cisco\\CCNx}
& \makecell{Fort.\\NSE}
& \makecell{Fort.\\ICS/SCADA}
& \makecell{ISA/IEC\\62443} \\
\midrule
\endfirsthead

\toprule
& \makecell{CyMetric\\80}
& \makecell{MMLU\\CompSec.}
& \makecell{Cisco\\CCNx}
& \makecell{Fort.\\NSE}
& \makecell{Fort.\\ICS/SCADA}
& \makecell{ISA/IEC\\62443} \\
\midrule
\endhead
\bottomrule
\endfoot

\bottomrule
\endlastfoot
Claude 3.5 Haiku & 96.2 & 85.0 & 76.5 & 47.5 & 81.4 & 66.0 \\
Claude 3.5 Sonnet & 95.0 & 87.3 & 86.5 & 59.6 & 83.7 & 81.1 \\
Claude 4.5 Haiku & 97.5 & 83.0 & 82.0 & 56.0 & 83.7 & 52.8 \\
Claude 4.5 Sonnet & 97.5 & 91.0 & 89.5 & 62.4 & 76.7 & 73.6 \\
DeepSeek v3 0324 & 95.0 & 84.0 & 82.0 & 62.4 & 79.1 & 77.4 \\
Gemini 2.5 Flash Lite & 95.0 & 78.0 & 74.0 & 52.5 & 81.4 & 64.2 \\
Gemini 2.5 Pro 0506 & 97.5 & 89.0 & 91.5 & 61.7 & 86.0 & 73.6 \\
Gemini Flash 2.0 & 97.5 & 82.0 & 80.8 & 56.7 & 79.1 & 74.3 \\
Gemma 2 27B & 93.8 & 79.0 & 66.0 & 50.4 & 76.7 & 43.4 \\
Gemma 2 9B & 91.2 & 74.0 & 54.5 & 42.6 & 74.4 & 35.8 \\
Gemma 3 12B & 92.5 & 83.0 & 65.0 & 42.6 & 69.8 & 43.4 \\
Gemma 3 27B & 96.2 & 81.0 & 68.0 & 45.4 & 67.4 & 37.7 \\
Gemma 3 4B & 85.0 & 68.0 & 51.0 & 36.9 & 53.5 & 39.6 \\
Llama2 70B & 86.2 & 73.0 & 51.5 & 44.0 & 65.1 & 54.7 \\
Llama3.1 405B & 100.0 & 84.0 & 80.6 & 53.9 & 80.9 & 63.0 \\
Llama3.1 70B & 92.5 & 82.6 & 77.4 & 51.8 & 80.0 & 64.2 \\
Llama3.1 8B & 92.5 & 81.0 & 54.5 & 41.8 & 65.1 & 54.7 \\
Llama3.2 3B & 83.8 & 68.0 & 53.0 & 41.8 & 74.4 & 37.7 \\
Llama4 Maverick & 93.8 & 89.0 & 76.5 & 56.0 & 79.1 & 66.0 \\
Llama4 Scout & 87.5 & 72.0 & 76.0 & 50.4 & 74.4 & 62.3 \\
Ministral 3B & 86.2 & 73.0 & 57.0 & 47.5 & 60.5 & 52.8 \\
Ministral 8B & 86.2 & 76.0 & 56.0 & 48.2 & 60.5 & 45.3 \\
Mistral 7B v0.3 & 88.8 & 72.0 & 52.0 & 45.4 & 62.8 & 60.4 \\
Mistral Large 2411 & 93.8 & 83.0 & 77.0 & 51.8 & 86.0 & 56.6 \\
Mistral Medium 3.1 & 97.5 & 82.0 & 84.5 & 55.3 & 81.4 & 67.9 \\
Mistral Small 3.1 24B & 93.8 & 82.0 & 79.5 & 53.2 & 88.4 & 66.0 \\
Mixtral 8x22B & 95.0 & 80.0 & 71.0 & 55.3 & 81.4 & 66.0 \\
Mixtral 8x7B & 92.5 & 77.0 & 61.5 & 46.1 & 60.5 & 62.3 \\
Moonshot AI Kimi K2 & 97.5 & 87.0 & 86.5 & 61.0 & 83.7 & 77.4 \\
OpenAI GPT 3.5 turbo & 95.0 & 80.0 & 67.0 & 51.8 & 81.4 & 66.0 \\
OpenAI GPT 4.1 & 96.2 & 86.0 & 84.0 & 58.2 & 86.0 & 69.8 \\
OpenAI GPT 4o & 96.2 & 85.0 & 85.8 & 58.2 & 86.5 & 68.3 \\
OpenAI GPT 5 & 97.5 & 88.0 & 85.5 & 53.2 & 88.4 & 69.8 \\
OpenAI GPT 5 Mini & 98.8 & 81.0 & 81.0 & 55.3 & 83.7 & 52.8 \\
OpenAI GPT 5 Nano & 92.5 & 76.0 & 64.0 & 40.4 & 74.4 & 32.1 \\
OpenAI GPT OSS 120B & 98.8 & 91.0 & 79.0 & 52.5 & 88.4 & 62.3 \\
OpenAI GPT OSS 20B & 97.5 & 85.0 & 73.0 & 53.9 & 74.4 & 60.4 \\
Qwen2 72B & 98.8 & 84.4 & 77.0 & 52.5 & 86.0 & 75.5 \\
Qwen2.5 72B & 97.5 & 86.6 & 78.9 & 53.9 & 87.4 & 59.2 \\
Qwen2.5 7B & 97.5 & 77.4 & 68.1 & 48.2 & 75.3 & 53.6 \\
Qwen3 14B & 93.8 & 87.0 & 74.0 & 53.0 & 80.6 & 71.1 \\
Qwen3 235B A22B & 91.2 & 89.0 & 80.5 & 51.8 & 79.1 & 73.6 \\
Qwen3 30B A3B & 97.5 & 83.0 & 67.0 & 51.1 & 81.4 & 66.0 \\
Qwen3 32B & 95.0 & 83.0 & 79.0 & 49.6 & 81.4 & 71.7 \\
Qwen3 8B & 95.0 & 79.0 & 67.0 & 50.4 & 67.4 & 60.4 \\
\end{longtable}

\end{document}